\documentclass[twocolumn,showpacs,preprintnumbers,amsmath,amssymb]{revtex4}

\usepackage{psfig} 
\usepackage{stmaryrd} 
\usepackage{graphicx} 
\usepackage{amssymb}  

\begin{document}

\title{Distributed Control by Lagrangian Steepest Descent}

   \author{David H. Wolpert$^1$}
   \author{Stefan Bieniawski$^2$}
   \author{Esfandiar Bandari$^2$}
   \affiliation{$^1$NASA Ames Research Center, Moffett Field, CA,
94035, and $^2$253 Durand, Dept. of Aeronautics
Stanford, CA 94305, stefanb@stanford.edu
    \tt \{dhw@ptolemy.arc.nasa.gov,stefanb@stanford.edu\}}


\begin{abstract}
Often adaptive, distributed control can be viewed as an iterated game
between independent players. The coupling between the players' mixed
strategies, arising as the system evolves from one instant to the
next, is determined by the system designer. Information theory tells
us that the most likely joint strategy of the players, given a value
of the expectation of the overall control objective function, is the
minimizer of a Lagrangian function of the joint strategy. So the goal
of the system designer is to speed evolution of the joint strategy to
that Lagrangian minimizing point, lower the expectated value of the
control objective function, and repeat. Here we elaborate the theory
of algorithms that do this using local descent procedures, and that
thereby achieve efficient, adaptive, distributed control.
\end{abstract}

\maketitle

\section{INTRODUCTION}

This paper considers the problem of adaptive distributed control
\cite{lamo93, woch96, meha99}. Typically in such problems, at each
time $t$ each control agent $i$ sets its state $x^t_i$ independently
of the other agents, by sampling an associated distribution,
$q^t_i(x^t_i)$.  Rather than directly via statistical dependencies of
the agents' states at the same time $t$, the coupling between the
agents arises indirectly, through the stochastic joint evolution of
their distributions \{$q^t_i$\} across time.

More formally, let time be discrete, where at the beginning of each
$t$ every agent sets its state (``makes its move''), and then the rest
of the system responds. Indicate the state of the entire system at
time $t$ as $z^t$. ($z^t$ includes $x^t$, as well as all stochastic
elements not being directly controlled.) So the joint distribution of
the moves of the agents at any moment $t$ is given by the product
distribution $q^t(x^t) = \prod_i q^t_i(x^t_i)$, and the state of the
entire system, given joint move $x^t$, is governed by $P(z^t \mid
x^t)$.

Now in general the observations by agent $i$ of aspects of the
system's state at times previous to $t$ will determine $q^t_i$. So
$q^t_i$ is statistically dependent on the previous states of the
entire system, $z^{\{t'<t\}}$.  In other words, the agents can be
viewed as players in a repeated game with Nature, each playing mixed
strategies $\{q^t_i\}$ at moment $t$
\cite{futi91,baol99,osru94,auha92,fule98}. Their interdependence
arises through information sets and the like, in the usual way. 

From this perspective what the designer of a distributed control
system can specify is the stochastic laws governing the updating of
the joint strategy.  In other words, the designer wishes to impose a
stochastic dynamics on a Multi-Agent System (MAS) that optimizes an
overall objective function of the state of the system in which the MAS
is embedded, $F(z)$.\footnote{Here we follow the convention that lower
$F$ is better. In addition, for simplicity we only consider objectives
that depend on the state of the system at a single instant; it is
straightforward to relax this restriction.} Formally, this means
inducing a joint strategy $q(x)$ with a good associated value of
$E_q(F) = \int dx q(x) E(F \mid x) \equiv \int dx q(x)
G(x)$.\footnote{For simplicity, here we indicate integrals of any
sort, including point sums for countable $x$, with the $\int$ symbol.}
Once such a $q$ is found, one can sample it to get a final $x$, and be
assured that, on average, the associated $F$ value is low. $G$ is
called the {\bf{world utility}}.

In this paper we elaborate a set of algorithms that iteratively update
$q^t$ in such a manner.  The algorithms presented here are based on
using steepest descent techniques to minimize a
$G$-parameterized Lagrangian, $L_G(q)$.\footnote{See
\cite{wolp04b,wolp04c,wolp03,biwo04c} for non-local techniques for
finding $q^t$, techniques that are related to ficticious play, and see
\cite{wole04} for techniques that exploit the Metropolis-Hastings
algorithm. Other non-local techniques are related to importance
sampling of integrals, and are briefly mentioned in \cite{wolp04a}.}
Because the descent is over Euclidean vectors $q$, these algorithms
can be applied whether the $x_i$ are categorical, continuous,
time-extended, or a mixture of the three. So in particular, they
provide a principled way to do ``gradient descent over categorical
variables''.

In the next section we first derive the Lagrangian $L_G(q)$ and
discuss some of its properties.

In the following section we show how to apply gradient descent (and
its embellishments) to optimization of the Lagrangian. If we view the
agents as engaged in a team game, all having the same utility $G$,
then this gradient descent is a distributed scheme for each agent to
update its strategy, in a way that will steer the game to a bounded
rational equilibrium \cite{wolp04a,wolp04b}.

In this section we also consider second order methods. In contrast to
gradient descent, in general any single application of Newton's method
to update a product distribution $q$ will result in a new distribution
$p(q)$ that is not a product distribution. So we must instead solve
for the product distribution $q'(p)$ having minimal Kullback-Leibler
distance to $p$. In this section we derive the rule for iterative
updating of our distribution so as to move $q$ in the direction of
$q'(p(q))$.

In practice any local descent scheme often requires Monte Carlo
sampling to estimate terms in the gradient.  To minimize the expected
quadratic error of the estimation, typically the game is changed from
being a team game. In other words, in general changing the agent's
utilities $g_i$ to not all equal $G$ will result in lower bias plus
variance of the estimation of the gradient, and therefore will speed
evolution to a good joint strategy. These and other techniques for
shrinking bias plus variance are discussed in \cite{wolp03,wolp04c}.

We end this section by mentioning some other techniques for improving the
Monte Carlo sampling. These include data-aging, and techniques for
managing the descent when it gets close to a border of the (simplex
of) allowed $q$, $\cal{Q}$. Most of these techniques introduced can be
used even with schemes for minimizing $L_G(q)$ other than gradient
descent.

In the final section we introduce some alternatives to $L_G(q)$,
designed to help speed convergence to a $q$ with low $E(G)$.
Miscellaneous proofs can be found in the appendix.

The general mathematical framework for casting control and
optimization problems in terms of minimizing Lagangians of probability
distributions is known as the theory of probability Lagangians. The
precise version where the probability distributions are product
distributions is known as ``Product Distribution'' (PD)
theory. \cite{wolp03}.  It has many deep connections to other fields,
including bounded rational game theory and statistical physics
\cite{wolp04a}.  As such it serves as a mathematical bridge connecting 
these disciplines. Some initial experimental results concerning the
use of PD theory for distributed optimization and distributed control
can be found in
\cite{mabi04,anbi04,lewo04,aiwo04,biwo04a,biwo04b}. See
\cite{wole04,wolp04c,wolp04d} for other uses and extensions of PD
theory.

\section{Product Distribution Lagrangians}

\subsection{The maxent Lagrangian}
Say the designer stipulates a particular desired value of $E(G)$,
$\gamma$. For simplicity, consider the case where the designer has no
other knowledge concerning the system besides $\gamma$ and the fact
that the joint strategy is a product distribution. Then information
theory tells us that the {\it {a priori}} most likely $q$ consistent
with that information is the one that maximizes entropy subject to
that information \cite{coth91,mack03,jabr03}.\footnote{In light of how
limited the information is here, the algorithms presented below are
best-suited to ``off the shelf" uses; incorporating more prior
knowledge allows the algorithms to be tailored more tightly to a
particular application.}  In other words, of all distributions that
agree with the designer's information, that distribution is the
``easiest'' one to induce by random search.

Given this, one can view the job of the designer of a distributed
control system as an iterative equilibration process. In the first
stage of each iteration the designer works to speed evolution of the
joint strategy to the $q$ with maximal entropy subject to a particular 
value of $\gamma$. Once we have found such a solution we can replace
the constraint --- replace the target value of $E(G)$ --- with a more
difficult one, and then repeat the process, with another evolution of
$q$ \cite{wolp03}.

Define the {\bf{maxent}} Lagrangian by
\begin{eqnarray}
L(q) &\equiv& \beta (E_q(G) - \gamma) - S \nonumber \\
&=& \beta (\int dx q(x) G(x) - \gamma) - S(q) ,
\label{eq:maxentlag}
\end{eqnarray}
\noindent where $S(q)$ is the Shannon entropy of $q$, $-\int dx q(x)
{\mbox{ln}}\frac{q(x)}{\mu(x)}$, and for simplicity we here take the
prior $\mu$ to be uniform.\footnote{Throughout this paper the terms in
any Lagrangian that restrict distributions to the unit simplices are
implicit. The other constraint needed for a Euclidean vector to be a
valid probability distribution is that none of its components are
negative. This will not need to be explicitly enforced in the
Lagrangian here.}. Writing it out, for a given $\gamma$, the
associated most likely joint strategy is given by the $q$ that
minimizes $L(q)$ over all those $(q, \beta)$ such that the Lagrange
parameter $\beta$ is at a critical point of $L$, i.e., such that
$\frac{\partial L}{\partial \beta} = 0$.

Solving, we find that the $q_i$ are related to each other via a set of
coupled Boltzmann equations (one for each agent $i$),
\begin{equation} 
q^\beta_i(x_i) \propto e^{-\beta E_{q^\beta_{(i)}}(G \mid x_i)}  
\label{eq:boltzG} 
\end{equation} 
\noindent where the overall proportionality constant for each $i$ is
set by normalization, the subscript $q^\beta_{(i)}$ on the expectation
value indicates that it is evaluated according to the distribution
$\prod_{j \ne i} q_j$, and $\beta$ is set to enforce the condition
$E_{q^\beta}(G) = \gamma$.  Following Nash, we can use Brouwer's fixed
point theorem to establish that for any fixed $\beta$, there must
exist at least one solution to this set of simultaneous equations. 

If we evaluate $E(G)$ at the solution $q^\beta$, we find that it is a
declining function of $\beta$. So in following the iterative procedure
of equilibrating and then lowering $\gamma$ we we will raise
$\beta$. Accordingly, we can avoid the steps of testing whether each
successive constraint $E(G) = \gamma$ is met, and simply monotonically
increase $\beta$ instead. This allows us to avoid ever explicitly
specifying the values of $\gamma$.

Simulated annealing is an example of doing this, where rather than
work directly with $q$, one works with random samples of it formed via
the Metropolis random walk algorithm \cite{kige83,
dilu93,cato98,vida93}. There is no {\it{a priori}} reason to use such
an inefficient means of manipulating $q$ however. Here we will work with
$q$ directly instead.  This will result in an algorithm that is not
simply ``probabilistic" in the sense that the updating of its
variables is stochastic (as in simulated annealing). Rather the very
entity being updated is a probability distribution.

\subsection{Shape of the maxent Lagrangian}

Consider $L$ as a function of $q$, with $\beta$ and $\gamma$ both
treated as fixed parameters.  (So in particular, $E_q(g)$ need not
equal $\gamma$.)  First, say that $q_{(i)}$ is also held fixed, with
only $q_i$ allowed to vary. This makes $E(g)$  linear in $q_i$. In
addition, entropy is a concave function, and the unit simplex is a
convex region. Accordingly, the Lagrangian of Eq.~\ref{eq:maxentlag}
has a unique local minimum over $q_i$.  So there is no issue of
choosing among multiple minima when all of $q_{(i)}$ is fixed. Nor is
there any problem of ``getting trapped in a local minimum'' in a
computational search for that minimum. Indeed, in this situation we
can just jump directly to that global optimum, via
Eq.~\ref{eq:boltzG}.

Now introduce the shorthand for any function $U(x)$,
\begin{eqnarray*}
[U]_{i,p}(x_i) \equiv \int dx_{(i)} U(x_i, x_{(i)})
p(x_{(i)} \mid x_i).
\end{eqnarray*}
\noindent So $[G]_{i,q_{(i)}}(x_i)$ is agent $i$'s ``effective'' cost
function, $E_{q_{(i)}}(G \mid x_i)$.  Consider the value
$E_{q^\beta_i}([G]_{i,q_{(i)}})$. This is the value of $E(G)$ at $i$'s
bounded rational equilibrium for the fixed $q_{(i)}$, i.e., it is the
value at the minimum over $q_i$ of $L$. View that value as a function
of $\beta$. One can show that this is a decreasing function.  In fact,
its derivative just equals the negative of the variance of
$[G]_{i,q_{(i)}}(x_i)$ evaluated under distribution $q^\beta_i(x_i)$
(see appendix). Combining this with the fact that $E(G)$ is bounded
below (for bounded $G$), establishes that the variance must go to zero
for large enough $\beta$. So as $\beta$ grows, $q^\beta_i(x_i)
\rightarrow 0$ for all $x_i$ that don't minimize $E_{q_{(i)}}(G \mid
x_i) $. In other words, in that limit, $q_i$ becomes Nash-optimal.

Next consider varying over all $q \in \cal{Q}$, the space of all
product distributions $q$.  This is a convex space; if $p \in \cal{Q}$
and $p' \in \cal{Q}$, then so is any distribution on the line
connecting $p$ and $p'$. However over this space, the $E(G)$ term in
$L$ is multilinear. So $L$ is not a simple convex function of $q$. So
we do not have  guarantees of a single local minimum.

The following lemma extends the technique of Lagrange parameters to
off-equilibrium points:

$ $

\noindent {\bf{Lemma 1:}} Consider the set of all vectors leading from
$x' \in {\mathbb{R}}^n$ that are, to first order, consistent with a
set of constraints over ${\mathbb{R}}^n$, \{$f_i(x) = 0$\}.  Of those
vectors, the one giving the steepest ascent of a function $V(x)$ is
$\vec{u} = \nabla V +
\sum_i \lambda_i \nabla f_i$, up to an overall proportionality
constant, where the $\lambda_i$ enforce the first order consistency
conditions, $\vec{u} \cdot \nabla f_i = 0 \;\; \forall i$.

$ $

Now examine the derivatives of $S(q)$ with respect to all components
of $q$, i.e., the $q$-gradient of the entropy. At the border of
$\cal{Q}$, at least one of the ln$(q_i)$ terms in those derivatives
will be negative infinite.  Combined with Lemma 1, this can be used to
establish that at the edge of $\cal{Q}$, the steepest descent
direction of any player's Lagrangian points into the interior of
$\cal{Q}$ (assuming finite $\beta$ and $\{G\}$). (This is reflected in
the equilibrium solutions Eq.~\ref{eq:boltzG}.) Accordingly, whereas
Nash equilibria can be on the edge of $\cal{Q}$ (e.g., for a pure
strategy Nash equilibrium), in bounded rational games any equilibrium
must lie in the interior of $\cal{Q}$. In other words, any equilibrium
(i.e., any local minimum) of a bounded rational game has non-zero
probability for all joint moves. So just as when only varying a single
$q_i$, we never have to consider extremal mixed strategies in
searching for equilibria over all $\cal{Q}$. We can use local descent
schemes instead
\cite{mabi04,biwo04a,wobi04}.

Lemma 1 can also be used to construct $G$ with more than one solution
to Eq.~\ref{eq:boltzG}. One can also show that for every player $i$ and
any point $q$ interior to $\cal{Q}$, there are directions in $\cal{Q}$
along which $i$'s Lagrangian is locally convex.  Accordingly, no
player's Lagrangian has a local maximum interior to $\cal{Q}$. So if
there are multiple local minima of $i$'s Lagrangian, they are
separated by saddle points across ridges.  In addition, the uniform
$q$ is a solution to the set of coupled equations Eq. \ref{eq:boltzG},
but typically is not a local minimum, and therefore must be a saddle
point.

Say that we were not restricting ourselves to product
distributions. So the Lagrangian becomes $L(p) = \beta (E_p(G) -
\gamma) - S(p)$, where $p$ can now be any distribution over $x$. There
is only one local minimum over $p$ of this Lagrangian, the
{\bf{canonical ensemble}}:
\begin{eqnarray*}
p^\beta(x) \propto e^{-\beta G(x)}
\end{eqnarray*}
\noindent In general $p^\beta$ is not a product distribution. However
we can ask what product distribution is closest to it.

Now in general, the proper way to approximate a target distribution
$p$ with a distribution from a subset $\cal{C}$ of the set of all
distributions is to first specify a misfit measure saying how well
each member of $\cal{C}$ approximates $p$, and then solve for the
member with the smallest misfit. This is just as true when $\cal{C}$
is the set of all product distributions as when it is any other set.

How best to measure distances between probability distributions is a
topic of ongoing controversy and research \cite{woma04a}. The most
common way to do so is with the infinite limit log likelihood of data
being generated by one distribution but misattributed to have come
from the other.  This is know as the {\bf{Kullback-Leibler
distance}} \cite{coth91,duha00,mack03}:
\begin{equation}
KL(p_1 \; || \; p_2) \equiv S(p_1 \; || \; p_2) - S(p_1)
\end{equation}
\noindent where $S(p_1 \; || \; p_2) \equiv -\int dx \; p_1(x)
{\mbox{ln}}[\frac{p_2(x)}{\mu(x)}]$ is known as the {\bf{cross
entropy}} from $p_1$ to $p_2$ (and as usual we implicitly choose
uniform $\mu$). The KL distance is always non-negative, and equals
zero iff its two arguments are identical. 

As shorthand, define the ``$pq$ distance'' as $KL(p \; || \; q)$, and
the ``$qp$ distance'' as $KL(q \; || \; p)$, where $p$ is our target
distribution and $q$ is a product distribution. Then it is
straightforward to show that the $qp$ distance from $q$ to target
distribution $p^{\beta}$ is just the maxent Lagrangian $L(q)$, up to
irrelevant overall constants. In other words, the $q$ minimizing the
maxent Lagrangian is $q$ with the minimal $qp$ distance to the
associated canonical ensemble.

However the $qp$ distance is the (infinite limit of the negative log
of) the likelihood that distribution $p$ would attribute to data
generated by distribution $q$. It can be argued that a better measure
of how well $q$ approximates $p$ would be based on the likelihood that
$q$ attributes to data generated by $p$. This is the $pq$ distance; it
gives a different Lagrangian from that of Eq.~\ref{eq:maxentlag}.

Evaluating, up to an overall additive constant (of the canonical
distribution's entropy), the $pq$ distance is
\begin{eqnarray*}
KL(p \; || \; q) = -\sum_i \int dx \; p(x) {\mbox{ln}}[q_i(x_i)].
\end{eqnarray*}
\noindent  This is equivalent to a  game where each
coordinate $i$ has the ``Lagrangian''
\begin{eqnarray}
L^*_{i}(q)
&\equiv& -\int dx_i \; p_i(x_i) {\mbox{ln}}[q_i(_i)] ,
\label{eq:marg}
\end{eqnarray}
\noindent where $p_i(x_i)$ is the marginal distribution $\int dx_{(i)}
p(x)$.
The minimizer of this is just $q_i = p_i \; \forall i$, i.e., each
$q_i$ is set to the associated marginal distribution of $p$. 

In the interests of space, the rest of this paper we restrict
attention to the $pq$ KL distance and associated maxent Lagrangian.

\section{Descent of the maxent Lagrangian}

\subsection{Gradient descent}

Consider the situation where each $x_i$ can take on a finite number of
possible values, $|{\xi}_i|$. Say we are iteratively evolving $q$ to
minimize $L$ for some fixed $\beta$, and are currently at some point
$q \in {\cal{Q}}$. Using Lemma 1, we can evaluate the direction from
$q$ within $\cal{Q}$ that, to first order, will result in the largest
drop in the value of $L(q)$:
\begin{eqnarray}
\frac{\partial L(q)}{\partial q_i(x_i = j)} &=&
u_i(j) - \sum_{x'_i} u_i(x'_i) / |{\xi}_i|,
\label{eq:grad}
\end{eqnarray}
\noindent where $u_i(j) \equiv \beta E(G \mid x_i = j) +
{\mbox{ln}}[q_i(j)]$. (Intuitively, the reason for subtracting
$\sum_{x'_i} u_i(x'_i) / |{\xi}_i|$ is to keep the distribution in the
set of all possible probability distributions over $x$, $\cal{P}$.)

Eq.~\ref{eq:grad} specifies the change that each agent should make to
its distribution to have them jointly implement a step in steepest
descent of the maxent Lagrangian. These updates are completely
distributed, in the sense that each agent's update at time $t$ is
independent of any other agents' update at that time.  Typically at
any $t$ each agent $i$ knows $q_i(t)$ exactly, and therefore knows
${\mbox{ln}}[q_i(j)]$. However often it will not know $G$ and/or the
$q_{(i)}$. In such cases it will not be able to evaluate the $E(G \mid
x_i = j)$ terms in Eq.~\ref{eq:grad} in closed form.

One way to circumvent this problem is to have those expectation values
be simultaneously estimated by all agents by repeated Monte Carlo
sampling of $q$ to produce a set of $(x, G(x))$ pairs. Those pairs can
then be used by each agent $i$ to estimate the values $E(G
\mid x_i = j)$, and therefore how it should update its distribution.
In the simplest version of such an update to $q$ only occurs once
every $L$ time-steps.  In this scheme only the samples $(x, G(x))$
formed within a block of $L$ successive time-steps are used at the end
of that block by the agents to update their distributions (according
to Eq.~\ref{eq:grad}).

\subsection{Higher order descent schemes}

In general, second order descent (e.g., Newton's method) of the maxent
Lagrangian is non-trivial, due to coupling that arises between the
agents and the requirement for associated matrix inversion. However
recall that one way to motivate the entropic product distribution
Lagrangian $L(q)$ starts by saying that what we really want is the
fully coupled canonical ensemble distribution, $p^\beta(x) \propto
exp(-\beta G(x))$. $L(q)$ then measures $qp$ KL-distance to that
desired distribution.  From this perspective, any given iteration of
second order descent of the maxent Lagrangian runs downhill on a
quadratic approximation to a distribution, a distribution that is
itself a product distribution approximation to the ultimate
distribution we want to minimize.

This suggests the alternative of making the approximations in the
opposite order.  In this approach we first make a quadratic
approximation (over the space of all $p$, not just all $q$) to the
maxent Lagrangian, $L(p)$. Via Newton's method this specifies a $p^*$
that minimizes that quadratic approximation. We can then find the
product distribution that is nearest (in $pq$ KL distance) to
$p^*$. This scheme is called {\bf{Nearest Newton}} descent.

The gradient and Hessian of $L(p)$ are given by

\begin{eqnarray*}
\frac{\partial{L}}{\partial p(x)}|_{p=p^0} &=& \beta G(x) + 1 +
     {\mbox{ln}}(p^0(x)) \\ \nonumber
\frac{\partial^2{L}}{\partial p(x) \partial p(x')}|_{p=p^0} &=&
\frac{\delta_{x,x'}}{p^0(x)} \\ \nonumber
\end{eqnarray*}

\noindent where $p^0$ is the current point. This Hessian is
positive-definite (given that the current $p$ is a member of
$\cal{P}$).  By simple Lagrange parameters, the general solution is
either on the border of $\cal{P}$, or if in the interior is given by

\begin{eqnarray*}
p^*(x) &=& -p^0(x) \; {\bigl{[}} \beta G(x) + {\mbox{ln}}(p(x)) +
  \lambda {\bigr{]}} \\ \nonumber
\end{eqnarray*}
\noindent where $\lambda$ is set by normalization. Solving, either $p^*$
is on the edge of the simplex, or
\begin{eqnarray*}
\frac{p^*(x)}{p^0(x)} &=& 1 -
S(p^0) - {\mbox{ln}}(p^0(x)) -
\beta [G(x) - E(G)] .  \\ \nonumber
\end{eqnarray*}

Note that the right-hand side is exactly the direction you should go
using (simplex-constrained) gradient descent of $L(p)$. So the
direction to $p^*$ from $p^0$ is given by the Hadamard product of
$p^0$ and the direction given by gradient descent. 

Now we can approximate $p^*$ with the product distribution having the
minimal KL distance to it. In particular, consider using $pq$ KL
distance rather than $qp$ KL distance. Recall that for this kind of KL
distance, the optimal product distribution approximation to a joint
distribution is given by the product of the marginals of that joint
distribution (see the discussion just below Eq.~\ref{eq:marg}). Say
that $p^0$ is in the form of a product distribution, $q^0$, i.e., that
we are starting from a product distribution. Then calculating the
marginals of the associated $p^*$ to get $q^*$ is trivial:

\begin{eqnarray}
\frac{q^*_i(j)}{q^0_i(j)} = 1 &-&
S(q^0_i) -
{\mbox{ln}}(q^0_i(j)) \nonumber  \\
&-& \beta [E(G \mid x_i = j) - E(G)] 
\label{eq:nnupdate}
\end{eqnarray}

Note that since the original quadratic approximation was over the full
joint space, this formula automatically takes into account inter-agent
couplings. In practice of course, it may make sense not to jump all
the way from $q^0$ to $q^*$, but only part-way there, to be
conservative. (In fact, if $q^*$ isn't in the interior of the simplex,
such partial jumping is necessary.) One potential guide to how far to
jump is the $pq$ KL distance from $p^*$ to $\prod_i q^*_i$. Unlike the
KL distances to the full joint Boltzmann distribution, we can readily
calculate this KL distance.

The conditional expectations in Nearest Newton are the same as those
in gradient descent. Accordingly, they too can be estimated via Monte
Carlo sampling, if need be. It's also worth noting that
Eq.~\ref{eq:nnupdate} has the same form as one would get by evaluating 
the Hessian of the maxent Lagrangian, so long as one ignored
inter-agent aspects of that Hessian.

\subsection{Practical issues}
\label{sec:prac}

In practice, the block-wise Monte Carlo sampling to estimate descent
directions described above can be prohibitively slow.  The estimates
typically have high variance, and therefore require large block size
$L$ to get a good descent direction. One set of ways to address this
is to replace the team game with a non-team game, i.e., for each agent 
$i$ have it estimate quantities $E(g_i \mid x_i = j)$ rather than $E(G 
\mid x_i = j)$, where each {\bf{private utility}} $g_i$ is chosen
to have much lower variance than $G$
\cite{wolp03,mabi04,wolp04c}.\footnote{Formally, this means that each
agent $i$ has a separate Lagrangian, formed from
Eq.~\ref{eq:maxentlag} by substituting $g_i$ for $G$. The associaed
joint solution $q$ is then given by substituting the appropriate $g_i$
for $G$ in each instance of the coupled equations Eq.~\ref{eq:boltzG}
(one instance for each $i$). See \cite{wolp04a} for the relation of
this to bounded rational game theory.}

Another useful technique is to allow samples from preceding blocks to
be re-used. One does this by first ``aging'' that data to reflect the
fact that it was formed under a different $q_{(i)}$ . For example, one
can replace the empirical average for the most recent block $k$,
\begin{eqnarray*}
\hat{G}_{i,j}(k) &\equiv&
\frac{\sum_{t = kL}^{kL + L} G(x^t) \delta_{x^t_i, j}}
{\sum_{t = kL}^{kL + L} \delta_{x^t_i, j}}
\end{eqnarray*}
\noindent with a weighted average over the expected $G$'s of all
preceding blocks,
\begin{eqnarray*}
\frac{\sum_m \hat{G}_{i,j}(m) e^{-\kappa (k - m)}}{\sum_m
e^{-\kappa (k - m)}}
\end{eqnarray*}
\noindent for some appropriate aging constant $\kappa$.\footnote{Not
all preceding $\hat{G}_{i,j}(m)$ need to be stored to implement this;
exponential ageing can be done online using 3 variables per $(i, j)$
pair.}

Typically such ageing allows $L$ to be vastly reduced, and therefore
the overall minimization of $L$ to be greatly sped up. For such small
$L$ though, it may be that the most recent block has {\it{no}} samples
of some move $x_i = j$. This would mean that $\hat{G}_{i,j}(k)$ is
undefined. One crude way to avoid such problems is to simply force a
set of samples of each such move if they don't occur of their own
accord, being careful to have the $x_{(i)}$ formed by sampling
$q_{(i)}$ when forming those forced samples.

Other useful techniques allow one to properly decrease the step size
as one nears the border of $\cal{Q}$.

\section{Other Lagrangians For Finding Minima Of $G$}

There are many alternative Lagrangians to the ones described
above. The section focuses on such alternative Lagrangians for the
purpose of finding argmin$_xG(x)$. Two classes of such Lagrangians are
investigated: variants of the Maxent Lagrangians, and variants of the
two types of KL-distance Lagrangians.

\subsection{Maxent Lagrangians}

Say that after finding the $q$ that minimizes the Lagrangian, we IID
sample that $q$, $K$ times. We then take the sample that has the
smallest $G$ value as our guess for the $x$ that minimizes $G(x)$. For
this to give a low $x$ we don't need the mean of the distribution
$q(G)$ to be low --- what we need is that the bottom tail of that
distribution is low. This suggests that in the $E(G)$ term of the
Maxent Lagrangian we replace
\begin{eqnarray*}
q(x) &\rightarrow& \\ 
&&q(x) \frac{ \Theta[ \kappa - \int dx' \;
q(x') \Theta[G(x) - G(x')]]} {\kappa}
\end{eqnarray*}
\noindent where $\Theta$ is the Heaviside theta function. 
This new multiplier of $G$ is still a probability distribution over
$x$. It equals 0 if $G(x)$ is in the worst $1 -
\kappa$ percentile (according to distribution $q$) of $G$ values, and
$\kappa^{-1}$ otherwise.  So under this replacement the $E(G)$ term in
the Lagrangian equals the average of $G$ restricted to that lower
$\kappa$'th percentile.  For $\kappa = K^{-1}$, our new Lagrangian
forces attention in setting $q$ on that outlier likely to come out of
the $K$-fold sampling of $q(G)$.

One can use gradient descent and Monte Carlo sampling to minimize this
Lagrangian, in the usual way. Note that the Monte Carlo process
includes sampling the probability distribution $\frac{ \Theta[ \kappa
- \int dx' \; q(x') \Theta[G(x) - G(x')]]}{\kappa}$ as well as the
$q_i$. This means that only those points in the best $\kappa$'th
percentile are kept, and used for all Monte Carlo estimates. This may
cause greater noise in the Monte Carlo sampling than would be the case
for $\kappa = 1$.

As an example, say that for agent $i$, all of its moves have the same
value of $E(G \mid x_i)$, and similarly for agent $j$, and say that
$G$ is optimal if agents $i$ and j both make move $0$. Then if we
modify the updating so that agent $i$ only considers the best values
that arose when it made move $0$, and similarly for agent $j$, then
both will be steered to prefer to make move 0 to their
alternatives. This will cause them to coordinate their moves in an
optimal manner. 

A similar modification is to replace $G$ with $f(G)$ in the maxent
Lagrangian, for some concave nowhere-decreasing function
$f(.)$. Intuitively, this will have the effect of coordinating the
updates of the separate $q_i$ at the end of the block, in a way to
help lower $G$. It does this by distorting $G$ to accentuate those
$x$'s with good values.  The price paid for this is that there will be
more variance in the values of $f(G)$ returned by the Monte Carlo
sampling than those of $G$, in general.

Note that if $q$ is a local minimum of the Lagrangian for $G$, in
general it will not be a local minimum for the Lagrangian of $f(G)$
(the gradient will no longer be zero under that replacement, in
general). So we can replace $G$ with $f(G)$ when we get stuck in a
local minimum, and then return to $G$ once $q$ gets away from that
local minimum. In this way we can break out of local minima, without
facing the penalty of extra variance. Of course, none of these
advantages in replacing $G$ with $f(G)$ hold for algorithms that
directly search for an $x$ giving a good $G(x)$ value; $x$ is a local
minimum of $G(x) \Leftrightarrow$ $x$ is a local minimum of $f(G(x))$.

An even simpler modification to the $E(G)$ term than those considered
above is to replace $G(x)$ with $\Theta[G(x) - K]$.  Under this
replacement the $E(G)$ term becomes the probability that $G(x) >
K$. So minimizing it will push $q$ to $x$ with lower $G$ values. For
this modified Lagrangian, the gradient descent update steps adds the
following to each $q_i(x_i)$:
\begin{eqnarray*}
\alpha\bigl[\beta q(G < K \mid
x_i) &+& {\mbox {ln}}(q_i(x_i)) \\ \nonumber
&-& \frac{\sum_{x'_i} \beta q(G < K \mid x'_i) +
{\mbox {ln}}(q_i(x'_i))}{\sum_{x'_i} 1}\bigr]. \nonumber
\end{eqnarray*}

In gradient descent of the Maxent Lagrangian we must Monte Carlo
estimate the expected value of a real number ($G$). In contrast, in
gradient descent of this modified Lagrangian we Monte Carlo estimate
the expected value of a single bit: whether $G$ exceeds
$K$. Accordingly, the noise in the Monte Carlo estimation for this
modified Lagrangian is usually far smaller.

In all these variants it may make sense to replace the Heaviside
function with a logistic function or an exponential. In addition, in
all of them the annealing schedule for $K$ can be set by periodically
searching for the $K$ that is (estimated to be) optimal, just as one
searches for optimal coordinate
systems \cite{wolp04a,wolp03}. Alternatively, a simple heuristic is to
have $K$ at the end of each block be set so that some pre-fixed
percentage of the sampled points in the block go into our calculation
of how to update $q$.

Yet another possibility is to replace $E(G)$ with the $\kappa$'th
percentile $G$ value, i.e., with the $K$ such that $\int dx' \; q(x')
\Theta(G(x') - K) = \kappa$. (To evaluate the partial derivative of
that $k$ with respect a particular $q_i(x_i)$ one must use implicit
differentiation.)

\subsection{KL-based Lagrangians}

Both the $qp$-KL Lagrangian and $pq$-KL Lagrangians discussed above
had the target distribution be a Boltzmann distribution over $G$. For
high enough $\beta$, such a distribution is peaked near argmin$_x
G(x)$. So sampling an accurate approximation to it should give an $x$
with low $G$, if $\beta$ is large enough.  This is why one way to
minimize $G$ is to iteratively find a $q$ that approximates the
Boltzmann distribution, for higher and higher $\beta$.

However there are other target distributions that are peaked about
minimizers of $G$. In particular, given any distribution $p'$, the
distribution
\begin{eqnarray*}
\theta_p(x) \equiv \frac{p(x) \Theta[K - G(x)]}{\int dx' \;
p(x') \Theta[K - G(x')]} \nonumber
\end{eqnarray*}
\noindent is guaranteed to be more peaked about such minimizers than
is $p$. So our minimization can be done by iterating the process of
finding the $q$ that best approximates $\theta_p$ and then setting $p
= q$. This is analogous to the minimization algorithm considered in
previous sections, which iterates the process of finding the $q$ that
best approximates the Boltzmann distribution and then increases
$\beta$.

For the choice of $pq$-KL distance as the approximation error, the
$q$ that best approximates $\theta_p$ is just the product of the
marginal distributions of $\theta_p$. So at the end of each iteration,
we replace 
\begin{eqnarray*}
q_i(x_i) &\leftarrow& \frac{\int dx'_{(i)} q'_{(i)}(x'_{(i)})
\Theta[K - G(x_i, x'_{(i)})]} {\int dx' \; q'(x')
\Theta[K - G(x')]} \\ \nonumber
&=& \frac{q'(G < K \mid x_i)} {q'(G < K)} \\ \nonumber
&=& \frac{q'(x_i \mid G < K)} {q'(x_i)} \nonumber
\end{eqnarray*}
\noindent where $q'$ is the product distribution being replaced. The
denominator term in this expression is known exactly to agent $i$, and
the numerator can be Monte-Carlo estimated by that agent using only
observed $G$ values. So like gradient descent on the Maxent
Lagrangian, this update rule is well-suited to a distributed
implementation.

Note that if we replace the Heaviside function in this algorithm with
an exponential with exponent $\beta$, the update rule becomes
\begin{eqnarray*}
q_i(x_i) \leftarrow
\frac{E(e^{-\beta G} \mid x_i)} {E(e^{-\beta G})}. \nonumber
\end{eqnarray*}
\noindent where both expectations are evaluated under $q'$, the
distribution that generated the Monte Carlo samples.  It's interesting
to compare this update rule with the parallel Brouwer update rule for
the team game \cite{biwo04a, wolp03, biwo04c}, to which it is very
similar.\footnote{That update is a variant of ficticious play, in
which we simultaneously replace each $q_i(x_i)$ with its ideal value
if $q_{(i)}$ were to be held fixed, given by Eq.~\ref{eq:boltzG}.} This
update is guaranteed to optimize the associated Lagrangian, unlike the
Brouwer update.  On the other hand, since it is based on the $pq$-KL
Lagrangian, as mentioned above there is no formal guarantee that this
alternative to Brouwer updating will decrease $E(G)$.

This update rule is also very similar to the adaptive importance
sampling of the original $pq$-KL approach discussed in
\cite{wolp03}. The difference is that in adaptive importance sampling
the $e^{-\beta G(x)}$ terms get replaced by $e^{-\beta G(x)} / q'(x)$.

Finally, consider using $qp$-KL distance to approximate $q'(x)
\frac{e^{\beta(K - G(x))}} {\int dx' \; q'(x') e^{\beta(K - G(x'))}}$
rather than $pq$-KL distance. In the Lagrangian for that distance the
$q'$ terms only contribute an overall additive constant. Aside from
that constant, this $qp$-KL Lagrangian is identical to the Maxent
Lagrangian.

\section{Conclusion}

Many problems in adaptive, distributed control can be cast as an
iterated game. The coupling between the mixed strategies of the
players arises as the system evolves from one instant to the next.
This is what the system designer determines.  Information theory tells
us that the most likely joint strategy of the players, given a value
of the expectation of the overall control objective function, is the
minimizer of a particular Lagrangian function of the joint
strategy. So the goal of the system designer is to speed evolution of
the joint strategy to that Lagrangian minimizing point, lower the
expectated value of the control objective function, and repeat. Here
we elaborate the theory of algorithms that do this using local descent
procedures, and that thereby achieve efficient, adaptive, distributed
control.

\section{Appendix}

This appendix provides proofs absent from the main text.

\subsection{Derivation of Lemma 1}

{\bf{Proof:}} Consider the set of $\vec{u}$ such that the directional
derivatives $D_{\vec{u}}f_i$ evaluated at $x'$ all equal 0. These are
the directions consistent with our constraints to first order. We need
to find the one of those $\vec{u}$ such that $D_{\vec{u}}V$ evaluated
at $x'$ is maximal.

To simplify the analysis we introduce the constraint that $|\vec{u}| =
1$. This means that the directional derivative $D_{\vec{u}}V$ for any
function $V$ is just $\vec{u} \cdot \nabla V$. We then use Lagrange
parameters to solve our problem. Our constraints on $\vec{u}$ are
$\sum_j u_j^2 = 1$ and $D_{\vec{u}}f_i(x') = \vec{u} \cdot \nabla
f_i(x') = 0 \; \; \forall i$.  Our objective function is
$D_{\vec{u}}V(x') = \vec{u} \cdot \nabla V(x')$.

Differentiating the Lagrangian gives
\begin{eqnarray*}
2 \lambda_0 u_i + \sum_i \lambda_i \nabla f = \nabla V \;\; \forall i.
\end{eqnarray*}
\noindent with solution 
\begin{eqnarray*}
u_i = \frac{\nabla V - \sum_i \lambda_i \nabla f}{2 \lambda_0}.
\end{eqnarray*}
\noindent $\lambda_0$ enforces our constraint on $|\vec{u}|$. Since we
are only interested in specifying $\vec{u}$ up to a proportionality
constant, we can set $2 \lambda_0 = 1$. Redefining the Lagrange
parameters by multiplying them by $-1$ then gives the result
claimed. {\bf QED.}

\subsection{Proof of claims following Lemma 1}

For generality, we provide the proofs in the general scenario where
the private utilities $g_i$ may differ from one another. See the
discussion in Section~\ref{sec:prac}.

\noindent {\bf{i)}} Define $f_i(q) \equiv \int dx_i q_i(x_i)$, i.e.,
$f_i$ is the constraint forcing $q_i$ to be normalized.  Now for any
$q$ that equals zero for some joint move there must be an $i$ and an
$x'_i$ such that $q_i(x'_i) = 0$.  Plugging into Lemma 1, we can
evaluate the component of the direction of steepest descent along the
direction of player $i$'s probability of making move $x'_i$:
\begin{eqnarray*}
&&\frac{\partial L}{\partial q_i(x_i)} + \lambda \frac{\partial
  f_i}{\partial q_i(x_i)} = \\
&&\;\;\;\; \beta E(g_i \mid x_i) + {\mbox{ln}}(q_i(x_i))  \\
&&\;\;\;\;\;\;\;\;\;\;\;\;\; - \frac{\int dx''_i [\beta
  E(g_i \mid x''_i) + {\mbox{ln}}(q_i(x''_i))]}{\int dx''_i 1}
\end{eqnarray*}

\noindent Since there must some $x''_i$ such that $q_i(x''_i) \ne 0$,
$\exists x_i$ such that $\beta E(g_i \mid x''_i) +
{\mbox{ln}}(q_i(x''_i))$ is finite.  Therefore our component is
negative infinite. So $L$ can be reduced by increasing
$q_i(x'_i)$. Accordingly, no $q$ having zero probability for some
joint move $x$ can be a minimum of $i$'s Lagrangian.

$ $

\noindent {\bf{ii)}} To construct a bounded rational game with
multiple equilibria, note that at any (necessarily interior) local
minimum $q$, for each $i$, 
\begin{eqnarray*}
&& \beta E(g_i \mid x_i) + {\mbox{ln}}(q_i(x_i)) = \\
&& \;\;\;\;\;\;\;\; \beta \int dx_{(i)} g_i(x_i,
x_{(i)}) \prod_{j\ne i} q_j(x_j) + {\mbox{ln}}(q_i(x_i))
\end{eqnarray*}
\noindent must be independent of $x_i$, by Lemma 1. So say there is a
component-by-component bijection $T(x) \equiv (T_1(x_1), T_2(x_2),
\ldots)$ that leaves all the $\{g_j\}$ unchanged, i.e., such that
$g_j(x) = g_j(T(x)) \; \forall x, j$ \footnote{As an example, consider
a congestion team game. In such a game all players have the same set
of possible moves, and the shared utility $G$ is a function only of
the $k$-indexed bit string $\{N(x, k)\}$, where $N(x, k) = 1$ iff
there is a move that is shared by exactly $k$ of the players when the
joint move is $x$. In this case $T$ just permutes the set of possible
moves in the same way for all players.}.

Define $q'$ by $q'(x) = q(T(x)) \; \forall x$. Then for any two values 
$x^1_i$ and $x^2_i$,
\begin{eqnarray*}
&&\beta E_{q'}(g_i \mid x^1_i) + {\mbox{ln}}(q'_i(x^1_i))  \\
&& \;\;\;\;\;\;\;\;\;\;\;\;\;- \;\beta E_{q'}(g_i \mid x^2_i) \;+\; {\mbox{ln}}(q'_i(x^2_i)) \\
&&\;\;\;\;\;\;\;\;\;\;\;\;\;\;\;\;\;\;\;\;\;\;= \\
&&\beta \int dx_{(i)}  g_i(x^1_i, x_{(i)}) \prod_{j\ne i} q_j(T(x_j)) \;+\;
		{\mbox{ln}}(q_i(T(x^1_i))) \\
&&\;\;\;\;\; - \; \beta \int dx_{(i)}  g_i(x^2_i, x_{(i)})
			\prod_{j\ne i} q_j(T(x_j))) \;+\;
		{\mbox{ln}}(q_i(T(x^2_i))) \\
&& \;\;\;\;\;\;\;\;\;\;\;\;\;\;\;\;\;\;\;\;\;\;= \\
&&\beta \int dx_{(i)}  g_i(x^1_i, T^{-1}(x_{(i)})) \prod_{j\ne i} q_j(x_j) \;+\;
		{\mbox{ln}}(q_i(T(x^1_i))) \\
&&\;\;\;\;\; - \; \beta \int dx_{(i)}  g_i(x^2_i, T^{-1}(x_{(i)}))
			\prod_{j\ne i} q_j(x_j)) \;+\;
		{\mbox{ln}}(q_i(T(x^2_i))) \\
&& \;\;\;\;\;\;\;\;\;\;\;\;\;\;\;\;\;\;\;\;\;\;= \\
&&\beta \int dx_{(i)}  g_i(T(x^1_i), x_{(i)})) \prod_{j\ne i} q_j(x_j) \;+\;
		{\mbox{ln}}(q_i(T(x^1_i))) \\
&&\;\;\;\;\; - \; \beta \int dx_{(i)}  g_i(T(x^2_i), x_{(i)}))
			\prod_{j\ne i} q_j(x_j)) \;+\;
		{\mbox{ln}}(q_i(T(x^2_i))) \\
&& \;\;\;\;\;\;\;\;\;\;\;\;\;\;\;\;\;\;\;\;\;\;= \\
&&\beta E_{q}(g_i \mid T(x^1_i)) \;+\; {\mbox{ln}}(q_i(T(x^1_i)))  \\
&& \;\;\;\;\;\;\;\;\;\;\;\;\;- \;\beta E_{q}(g_i \mid T(x^2_i))
	\;+\; {\mbox{ln}}(q_i(T(x^2_i))) 
\end{eqnarray*}
where the invariance of $g_i$ was used in the penultimate step. Since
$q$ is a local minimum though, this last difference must equal
0. Therefore $q'$ is also a local minimum.

Now choose the game so that $\forall i, x_i, T(x_i) \ne x_i$. (Our
congestion game example has this property.) Then the only way the
transformation $q \rightarrow q(T)$ can avoiding producing a new
product distribution is if $q_i(x_i) = q_i(x'_i) \; \forall i, x_i,
x'_i$, i.e., $q$ is uniform. Say the Hessians of the players'
Lagrangians are not all positive definite at the uniform $q$. (For
example have our congestion game be biased away from uniform
multiplicities.) Then that $q$ is not a local minimum of the
Lagrangians. Therefore at a local minimum, $q \ne q(T)$. Accordingly,
$q$ and $q(T)$ are two distinct equilibria.

$ $

\noindent {\bf{iii)}}  To establish that at any $q$ there is always a
direction along which any player's Lagrangian is locally convex, fix
all but two of the $\{q_i\}$, $q_0$ and $q_1$, and fix both $q_0$ and
$q_1$ for all but two of their respective possible values, which we
can write as $q_0(0), q_0(1), q_1(0)$, and $q_1(1)$, respectively. So
we can parameterize the set of $q$ we're considering by two real
numbers, $x \equiv q_0(0)$ and $y \equiv q_1(0)$. The $2 \times 2$
Hessian of $L$ as a function of $x$ and $y$ is
\[ \left( \begin{array}{cc}
\frac{1}{x} + \frac{1}{a-x} & \alpha  \\
\alpha & \frac{1}{y} + \frac{1}{b-y}  \end{array} \right)\]
\noindent where $a \equiv 1 - q_0(0) - q_0(1)$ and $b \equiv 1 -
q_1(0) - q_1(1)$, and $\alpha$ is a function of $g_i$ and $\prod_{j
\ne 0,1} q_j$. Defining $s \equiv \frac{1}{x} + \frac{1}{a-x}$ and 
$t \equiv \frac{1}{y} + \frac{1}{b-y}$, the eigenvalues of that
Hessian are 
\begin{eqnarray*}
\frac{s + t \pm \sqrt{4 \alpha^2 + (s - t)^2}}{2} .
\end{eqnarray*}
The eigenvalue for the positive root is necessarily
positive. Therefore along the corresponding eigenvector, $L$ is
convex at $q$. {\bf{QED.}}

$ $

\noindent {\bf{iv)}} There are several ways to show that the value
of $E_{q^\beta_i}([g_i]_{i,q_{(i)}})$ must shrink as $\beta$ grows. Here
we do so by evaluating the associated derivative with respect to
$\beta$.

Define $N(U) \equiv \int dy \; e^{-U(y)}$, the normalization constant
for the distribution proportional to $e^{-U(y)}$. View the
$x_i$-indexed vector $q^\beta_i$ as a function of $\beta, g_i$ and
$q_{(i)}$. So we can somewhat inelegantly write $E(g_i) =
E_{q^\beta_i(\beta, g_i, q_{(i)}), q_{(i)}}(g_i)$. Then one can expand
\begin{eqnarray*}
\frac{\partial E(g_i)}{\partial \beta} &=&
-\frac{\partial^2 {\mbox{ln}} (N(\beta [g_i]_{i,q_{(i)}}))}{\partial \beta^2} \\
&=& -{\mbox{Var}}( [g_i]_{i,q_{(i)}} )
\end{eqnarray*}
\noindent where the variance is over possible $x_i$, sampled according
to $q^\beta_i(x_i)$. {\bf{QED.}}

\bibliographystyle{IEEEtran}


\end{document}